\documentclass[12pt]{iopart}
\usepackage{times}
\usepackage{graphicx}

\begin{document}

\title[Mapping of few-electron wave-functions in semiconductor
nanocrystals]{Mapping of few-electron wave-functions in semiconductor
nanocrystals - evidence of exchange interaction}
\author{Michael Tews and Daniela Pfannkuche}
\address{I. Institute of Theoretical Physics, University of Hamburg,
Jungiusstr. 9, 20355 Hamburg, Germany}
\begin{abstract}
  The influence of the tip-substrate bias induced electric field in a
  scanning-tunneling-spectroscopy experiment on charged InAs nanocrystals
  is studied. Calculating the ground and first excited many-particle
  state for five electrons occupying the quantum dot reveals a Stark-induced
  reordering of states by increasing the electric field strength. It is
  shown that this reordering of states is accompanied by a symmetry change
  of the local density of states (LDOS), which in principal is observable
  in a wave-function mapping experiment. Since in the usually performed 
  experiments the electric field can not be directly controlled, we investigate
  the crystal size dependence of the 5-electron LDOS symmetry. It is found that
  the symmetry changes from spherical to torus-like by increasing the 
  nanocrystal radius.
\end{abstract}

\section{Introduction}
  Wave-function mapping in semiconductor quantum dots (QD) has
  recently attracted much interest since it serves as the ultimate tool to
  study the electronic structure of those dots
  \cite{millo:06:2001,grandidier:07:2000,vdovin:10:2000}.
  Knowing the actual shape of the electronic densities contributes to a
  better understanding of the QD's electronic structure. This knowledge is
  crucial with respect to the possible importance of semiconductor
  QDs as the ultimate building blocks of optoelectronic and nanoelectronic
  devices.

  Next to various experimental techniques available for different dot types,
  recent scanning-tunneling-microscopy (STM) measurements also allow an
  imaging of electronic densities in colloidal nanocrystals \cite{millo:06:2001}.
  Those nanocrystals are nearly spherical in shape resulting in atomic-like
  symmetries and degeneracies of the electronic states
  \cite{banin:08:1999,alperson:09:1999}. Nevertheless certain experiments
  \cite{millo:06:2001} show densities with a torus-like symmetry.
  In the regime where electrons tunnel through neutral nanocrystals we could
  attribute this broken symmetry previously \cite{tews:02:2002}
  to the electric field induced by the applied STM voltage. Although 
  scanning-tunneling-spectroscopy (STS) experiments on colloidal nanocrystals
  are possible in this transport regime \cite{katz:05:2001}, in the usually
  performed experiments the nanocrystal gets charged by increasing the STM 
  voltage. Therefore we study in this work the effect of the STM voltage on the
  electronic states in the regime of tunneling through already charged 
  nanocrystals. Nevertheless the results of the earlier published single-particle
  calculation are also of great importance for this regime and are therefore
  reviewed in the first part of this paper.

  We present a calculation of the many-particle conduction band (CB) 
  states within a particle-in-a-sphere model taking into account the 
  electric field due to the applied STM voltage. We mainly concentrate
  in this work on a Stark effect induced reordering of states for five 
  electrons occupying the nanocrystal. It will be shown that this 
  reordering corresponds to a change of the LDOS symmetry which in 
  principal is observable in a wave-function mapping experiment.

\section{Model}
  \label{sec:model}
\begin{figure}[h]
    \begin{center}
      \includegraphics[scale=0.20]{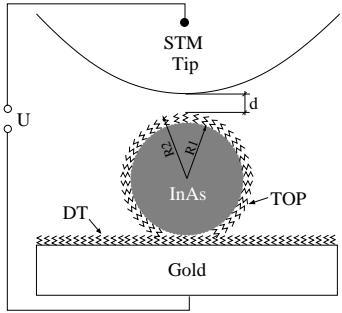}
    \end{center}
    \caption{Scanning tunneling spectroscopy of a single InAs nanocrystal.
             The InAs nanocrystal with a typical radius of a few
             nanometers are linked to a gold substrate by hexane dithiol
             molecules (DT). Trioctylphosphin (TOP) molecules form
             a ligand shell around the nanocrystal. At 4.2K the tunnel
             current is measured as a function of the applied voltage U
             between tip and substrate.}
    \label{fig_experiment}
  \end{figure}

  A sketch of the experimental setup in a STS experiment on colloidal
  InAs nanocrystal quantum dots is shown in Fig. \ref{fig_experiment}. To
  obtain a tunnel spectrum the
  tip is positioned above a single nanocrystal attached to the substrate
  via hexane dithiol molecules. Keeping the tip-crystal distance constant
  the differential conductance as a function of applied voltage shows
  sharp peaks \cite{banin:08:1999,millo:06:2000,bakkers:09:2000}.
  For a detailed understanding of the experimental data it is thus
  crucial to know the discrete energy spectrum of the QD since the
  obtained peak positions $U_{peak}$ are directly related to the
  electronic dot structure \cite{bakkers:09:2000}:
  \begin{eqnarray}
    eU_{peak}(N,N+1) = \gamma [E(N+1,\alpha)-E(N,\beta)] \label{Eq_peakpos}\\
    E(N,\alpha=\{n_i\}) = \sum_{i} n_i \epsilon_i + V_Q^{tot}(\{n_i\})
    \label{Eq_TotalEnergy}
  \end{eqnarray}
  where the pre-factor $\gamma$ depends on the capacitive electrostatic
  geometry. For a very asymmetric tip-dot dot-substrate capacity
  distribution $\gamma$ is close to one \cite{bakkers:09:2000}.
  The total energy $E(N,\alpha=\{n_i\})$ of excitation level $\alpha$ 
  with $N=\sum_i n_i$ electrons,
  where $n_i$ denotes the occupation number of state $i$, is written as a
  sum of occupied single particle levels with energies $\epsilon_i$ and
  the total charging energy $V_Q^{tot}$. Hereby $V_Q^{tot}$ includes both,
  the direct Coulomb interaction and all correlations.

  In order to obtain the discretised energy levels (\ref{Eq_TotalEnergy})
  of the CB electrons we use a single-band envelope wave-function
  approximation. The confinement due to the finite crystal size is modeled
  by a spherical potential well with finite depth \cite{brus:05:1984}.
  Depending on the crystal radius and the material constants of the used
  semiconductor those QDs can form a quite strong confinement. This
  confinement can lead to a size quantization of the electronic energies
  in the order of the semiconductor gap. Hence
  non-parabolicity effects of the CB have to be taken into account.
  Accounting for this effect we use an energy dependent effective mass
  approach \cite{burt:03:1992} with
  \begin{equation}
    m^*(E) = m^*(0) [1+{E}/{E_g}] \label{eqn:energymass}
  \end{equation}
  where $m^*(0)$ is the bottom CB effective mass (0.0239 $m_e$ in
  InAs) and $E_g$ the bulk
  energy gap. This approach has proven to by quite successful in
  reproducing the energy gap between the single particle
  ground and first excited state in InAs nanocrystals \cite{tews:02:2002}.
  In the STS setup of Fig. \ref{fig_experiment} with typical
  voltages up to $U \approx 2V$ \cite{millo:06:2001,banin:08:1999}
  applied on a tip-substrate distance of a few nanometers, the QD is
  exposed to a considerable electric field. Therefore the model 
  Hamiltonian for a single CB electron reads 
  \begin{equation}
    H = -\frac{\hbar^2}{2m^*(E)}\nabla^2+V_0 \Theta(r-R_1)
    -e\Phi_e(\vec{r})
    \label{eq_hamiltonian}
  \end{equation}
  with $\Phi_e(\vec{r})$ being the electrostatic potential. The 
  confinement potential is described by a Heaviside function
  $\Theta(r-R_1)$ with the potential well depth $V_0$.

\begin{figure}[h]
    \begin{center}
      \includegraphics[scale=0.45]{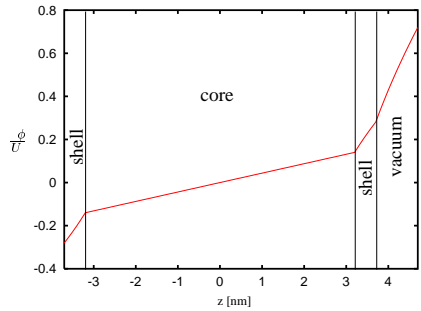}
    \end{center}
    \caption{Electrostatic potential between STM tip and gold substrate
             divided by the applied tip-substrate voltage $U$ plotted
             along the tip direction. The potential is calculated for
             an InAs nanocrystal of 3.2nm in radius and a tip-crystal
             distance of 1nm. The substrate is on the left and the tip
             starts on the very right. The dielectric constants 
             $\epsilon_1=15.15$ \cite{madelung:1996} and $\epsilon_2 =2.1$
             \cite{kim:06:2001} have been used for the InAs dot and
             the ligand shell, respectively.}
    \label{fig_potentialdrop}
  \end{figure}

  The electrostatic potential $\Phi_e(\vec{r})$ is obtained from a realistic
  modeling of the electrostatic environment in the experimental setup
  of Fig. \ref{fig_experiment}.
  While a STM tip has to terminate in a single atom in order to achieve
  atomic resolution, the macroscopic tip size is usually about one order in
  magnitude bigger than the here studied nanocrystals \cite{wiesendanger:1994}.
  Other than a macroscopic metallic tip, a single terminating atom is not
  able to substantially focus the electric field. Over the nanocrystal
  size, we can therefore assume the field between tip and substrate to be
  homogeneous in the absence of the QD. Since these nanocrystals are
  typically surrounded by ligands with a quite different relative
  dielectric constant compared to the semiconducting crystal we model
  the QD as a jacketed dielectric sphere. Extending the textbook
  calculation of a dielectric sphere placed in a homogeneous electric
  field ${\cal E}_{hom}$ \cite{griffiths_edyn:1998} to such a structure
  leads to a potential inside the QD of
  \begin{eqnarray}
    \Phi_{e}(r,\theta) =
    & & \frac{9\epsilon_{2} {\cal E}_{hom} r cos\theta}
    {2\epsilon_{2}^2+\epsilon_{1} \epsilon_{2}+4\epsilon_{2}+
    2\epsilon_{1}+2(\frac{R_{1}}{R_{2}})^3[\epsilon_{1}\epsilon_{2}
     +\epsilon_{2}-\epsilon_{1} - \epsilon_{2}^2]}
    \label{eq_corePotential}
  \end{eqnarray}
  with the relative dielectric constants $\epsilon_1$ and $\epsilon_2$ of the
  nanocrystal and the ligand shell, respectively. As in the case of a dielectric
  sphere without a shell \cite{griffiths_edyn:1998} the core potential
  $\Phi_{e}$ is still the potential of a homogeneous field. The
  field ${\cal E}_{hom}$  occurring in (\ref{eq_corePotential}) is
  not directly accessible. It is obtained by equating the voltage $U$
  with the potential drop between tip and substrate of the
  inhomogeneous field outside the QD. Knowing the electrostatic potential
  inside and also outside the nanocrystal the voltage drop over 
  both tunneling barriers can be obtained by plotting the potential along
  z-direction as shown in Fig. \ref{fig_potentialdrop}.
  The pre-factor $\gamma$ of equation (\ref{Eq_peakpos}) can also be extracted
  from the electrostatic potential by simply relating the potential drop
  between tip and QD-center to the drop between QD-center and substrate.

  For nanocrystals charged by $N$ CB electrons Coulomb interaction between
  those electrons will be present such that the many particle Hamiltonian
  can be written as
  \begin{equation}
    H = -\sum_{i=1}^N \left[ \frac{\hbar^2}{2 m^*}\nabla_i^2-V_0 \Theta(r_i-R_1)
        +e\Phi_e(\vec{r}_i)\right]+\sum_{i>j}^N\frac{e^2}{4 \pi \epsilon_0
        \epsilon_1r_{ij}}
        \label{eq_NHamiltonian}
  \end{equation}
  with $\epsilon_1$ being the dielectric constant of the semiconducting
  nanocrystal. Due to the finite potential well depth the electronic 
  wave-functions will to some extend leak out into the ligand shell. The
  much smaller dielectric constant in the ligand shell 
  leads to an enhanced Coulomb interaction in this outer region compared to 
  inside the QD. Therefore the Coulomb energy will be underestimated
  by this model Hamiltonian. In fact we found the charging energy for
  InAs nanocrystals about a factor two smaller than the experimentally
  observed values \cite{banin:08:1999}. To account for this effect one has to
  replace the Coulomb operator in (\ref{eq_NHamiltonian}) by the proper 
  Green's function of a dielectric sphere \cite{goldoni:06:2000,hallam:01:1996}.
  Instead we use  in this work $\epsilon_1$ as a fitting 
  parameter to match the experimentally found charging energy. 
  Mirror charges induced in the metallic tip and substrate are also 
  neglected.

\section{Single-particle calculations}
\label{sec:singleparticle}
\begin{figure}[h]
    \begin{flushright}
\includegraphics[scale=0.7]{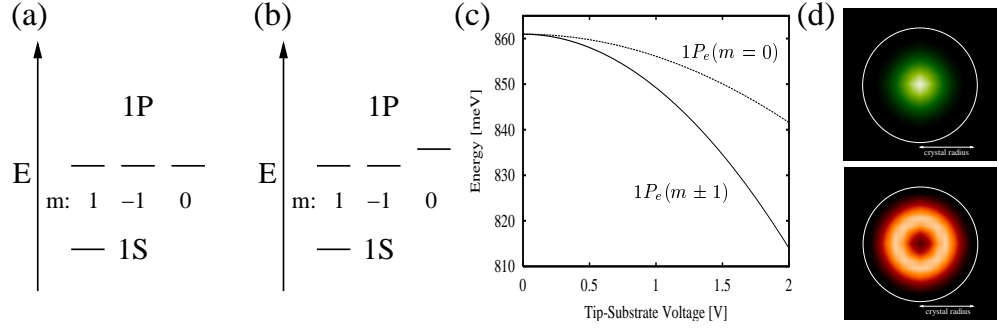}
    \end{flushright}
    \caption{a) Sketch of ground and first excited state
             without electric field. b) Single particle energy levels
             with an electric field applied along z-direction. The $1P_e$
             degeneracy is partly lifted. c) Stark splitting of the $1P_e$ level
             in a 3.2 nm radius InAs nanocrystal as a function of the applied
             tip-substrate voltage. For this calculation a confining potential
             well depth of $V_0=3$ eV has been used \cite{williamson:06:1999}. 
             d) Electronic density of state $1P_e(m=0)$ (top) and $1P_e(m\pm1)$
             (bottom) viewed along the applied electric field.} 
    \label{fig_starkeffect}
  \end{figure}

  The single-particle Hamiltonian (\ref{eq_hamiltonian}) without
  electrostatic potential separates in an angular and a radial
  part where the angular part is solved by spherical harmonics. The
  radial Schr\"odinger equation is solved by spherical Bessel functions
  $j_l$ inside the well and spherical Hankel functions $h_l$ outside
  \cite{schiff:1993}. The continuity conditions at the potential step
  lead to a set of transcendental equations determining the energy
  levels
  \begin{eqnarray}
    \lefteqn{ \alpha h_l(i \beta R_1) \left[ l j_{l-1}(\alpha R_1)
    -(l+1)j_{l+1}(\alpha R_1) \right] } \nonumber \\ &=&
     i \beta j_{l}(\alpha R_1) \left[ l h_{l-1}(i\beta R_1)
    -(l+1)h_{l+1}(i \beta R_1) \right] \label{eqn:trans}
  \end{eqnarray}
  with $\alpha = \sqrt{2m^* E}/\hbar$ and $\beta=\sqrt{2m^*(V-E)}/\hbar$.
  For small nanocrystals where the energy levels are in the order of
  the bulk energy gap the energy dependent mass of equation (\ref{eqn:energymass})
  can be directly inserted into the transcendental equation (\ref{eqn:trans}).
  Like in the hydrogen atom the single-particle ground-state has a s-type
  wave-function (hereafter referred to the $1S_e$ state, where the
  subscript $e$ denotes an electron rather than a hole state). Other
  than in hydrogen the allowed orbital quantum numbers are not restricted
  by the principal quantum number. Hence the first excited state has the
  quantum numbers $n=1$ and $l=1$ which we will call the $1P_e$ state.
  Owing to the spherical harmonics, this state is threefold degenerate
  in the three magnetic quantum numbers $m$=-1, 0 and 1 as shown in
  Fig. \ref{fig_starkeffect}a.

  Knowing the single-particle states without an electric field allows
  now to calculate the Stark effect either by an exact diagonalization
  or by perturbation theory. In the case of InAs nanocrystals the results
  of a calculation in second order perturbation theory has proven to
  be sufficient \cite{tews:02:2002}. Although the $1S_e$ state also
  shows a Stark effect, we concentrate here on the first excited $1P_e$
  state. In contrast to hydrogen, the first excited state does not show a linear
  Stark effect, owing to the lack of s-p degeneracy in such step-like
  spherical potential. The $1P_e$ degeneracy is lifted by the
  quadratic Stark effect, such that the energy of the $1P_e(m\pm1)$
  wave-functions oriented perpendicular to the electric field are
  lowered compared to the $1P_e(m=0)$ wave-function oriented along
  the field (see Fig. \ref{fig_starkeffect}b).

  Using the material parameters of a 3.2 nm InAs crystal leads to
  a Stark splitting of about 15 meV (see Fig. \ref{fig_starkeffect}c)
  at an applied voltage of 1.4 V which corresponds to the experimental
  situation in \cite{millo:06:2001}. This Stark-induced energy
  splitting is experimentally resolvable and
  allows at an appropriate voltage tunneling into the energetically
  lower $1P_e(m\pm1)$ states without tunneling through the $1P_e(m=0)$
  state. The qualitatively different density distributions of those
  split states (see Fig. \ref{fig_starkeffect}d), namely torus-like
  for the $1P_e(m\pm1)$ and spherical for the superposition of all 
  three p-type orbitals, are observed in a STM experiment by Millo 
  {\it et al.} \cite{millo:06:2001}. Therefore, the obtained 
  Stark-induced degeneracy lifting of the first excited $1P_e$ state
  serves as a possible explanation for the experimentally observed
  mapping of the $1P_e(m\pm1)$ wave-functions only.

\section{Many-particle calculation}
  In the usually performed experiments the p-type orbitals are
  available for tunneling only if the nanocrystal is already occupied
  by at least two electrons. Coulomb interaction between the CB
  electrons might be important and therefore the question arises
  if the single particle results of last section concerning the
  LDOS symmetry are still valid in this transport regime.

  Solutions of the many-particle Schr\"odinger equation belonging to Hamiltonian
  (\ref{eq_NHamiltonian}) were found by an exact diagonalization procedure in 
  the basis of wave-functions solving the many-particle Schr\"odinger equation 
  without Coulomb interaction and without electrostatic potential. 
  In order to keep the obtained matrices as small as possible we used the
  $\hat{L}_z$ and $\hat{S}_z$ symmetry of Hamiltonian
  (\ref{eq_NHamiltonian}) by selecting the needed basis functions. Furthermore
  the used basis is terminated by an energy cutoff, meaning that only
  Slater determinants with an energy below some threshold
  are used. The accuracy of a calculated energy level is estimated by the
  relative difference to the energy obtained by using a basis of just half
  the size. For all many-particle calculations presented in this work the
  relative error in energy is smaller than $1.5 \cdot 10^{-3}$.
  
  \subsection{Channel with 3 and 4 electrons}
\begin{figure}[h]
    \begin{center}
      \includegraphics[scale=0.5]{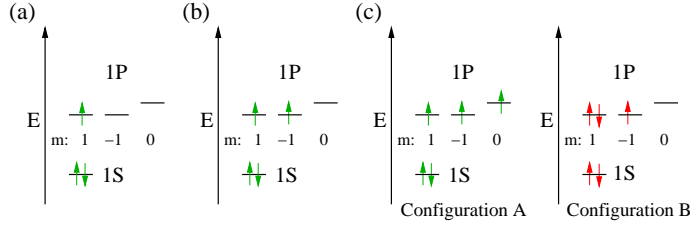}
    \end{center}
    \caption{(a) Leading ground-state configuration for three electrons. (b)
             Ground-state configuration according to Hund's rule for
             4 electrons occupying the nanocrystal. (c) Two possible
             ground-state configurations for five electrons.}
    \label{fig_configuration}
  \end{figure}

  In the case of the first p-type channel a third electron does tunnel
  through the QD which is already occupied by a full $1S_e$
  shell. Owing to the Stark-induced degeneracy lifting the third
  electron will tunnel through the energetically lower $1P_e(m\pm1)$
  state. Therefore the leading term of the 3-electron ground-state corresponds
  to the configuration shown in Fig. \ref{fig_configuration}a. Due to the small 
  crystal size and the high dielectric constant ($\epsilon_{InAs}$ = 15.15) in 
  those semiconducting nanocrystals the Coulomb energy is usually small compared
  to the kinetic energy (about 100 meV compared to 320 meV
  in a 3.2 nm InAs dot \cite{banin:08:1999}). Therefore the
  LDOS will be dominated by this leading configuration and
  hence torus-like in symmetry.

  Since the $1P_e(m\pm1)$ states are still two-fold degenerate the fourth 
  electron will also tunnel through one of those energetically lower states. 
  In order to gain exchange energy both p-electrons will align their spins
  (Hund's rule) which is the configuration shown in Fig. 
  \ref{fig_configuration}b. The LDOS symmetry of this 4-electron ground-state 
  will again be torus-like.

  \subsection{Channel with 5 electrons}
\begin{figure}[h]
    \begin{center}
      \includegraphics[scale=0.5]{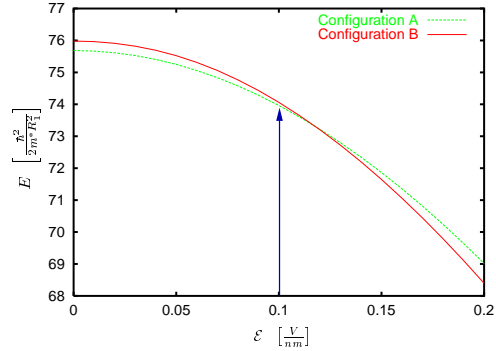}
    \end{center}
    \caption{Ground and first excited state energies of five
             electrons versus the electric field strength inside a 3.2 nm 
             radius InAs nanocrystal. The found ground-state crossing corresponds
             to a change in the LDOS symmetry. For this 3.2 nm dot the 
             5-electron channel is available at a tip-substrate voltage of about
             1.5 V leading to an electric field strength indicated by the vertical
             arrow.}
    \label{fig_GSCrossing}
  \end{figure}

  A more interesting situation arises for the fifth electron tunneling
  through the nanocrystal, where a competition between Stark energy
  on one hand and exchange energy on the other hand arises. Depending on
  how strong the splitting of the $1P_e$ states is, two different
  ground-states are possible. For a small splitting the configuration with all 
  three p-type orbitals, $1P_e(m=\pm1$) and  $1P_e(m=0)$, occupied each by one
  electron with their spins aligned will be favored (see configuration A in Fig.
  \ref{fig_configuration}c). In this configuration the fifth electron
  has to pay some Stark energy but gains exchange energy.  On the other hand
  if the splitting becomes too big it is energetically more favorable for
  the fifth electron to occupy also a $1P_e(m\pm1)$ state, thereby saving Stark
  energy. Due to the necessary spin flip however, it has to pay exchange energy
  (see configuration B in Fig. \ref{fig_configuration}c).
  This competition between Stark and exchange energy leads to a ground-state
  crossing with increasing electric field strength (see Fig.
  \ref{fig_GSCrossing}).
  The interesting point is that this crossing also corresponds to a change
  in the LDOS symmetry from spherical to torus-like by increasing
  the electric field. In contrast to the 3- and 4-electron channel
  the LDOS symmetry of the 5-electron channel can be controlled by the
  electric field applied to the nanocrystal.

  Unfortunately the electric field strength applied to the nanocrystal is
  determined by the corresponding peak position found in a STS experiment and,
  therefore, it is experimentally not straight forward to switch the LDOS
  symmetry forth and back between torus-like and spherical in a wave-function
  mapping experiment. For a 3.2 nm InAs nanocrystal the 5-electron channel is
  available at a tip-substrate voltage of about 1.5 V leading to an electric
  field of about 0.1 V/nm. As indicated by the vertical arrow in Fig. 
  \ref{fig_GSCrossing} the LDOS symmetry of the ground-state is at this  
  field strength still spherical.
  On the other hand the STS-peak positions depend on
  the QD radius such that we now stress the question how the LDOS symmetry
  of the 5-electron ground-state changes with the nanocrystal radius.

  In order to answer the question how the ground-state symmetry depends
  on the crystal size the scaling behavior of the Coulomb operator versus
  the electrostatic potential in (\ref{eq_NHamiltonian}) with respect
  to the dot radius is studied. 
  Whereas it is clear that the Coulomb operator scales with $R_1^{-1}$
  the scaling of the electrostatic potential is not easily foreseen. As
  shown in equation (\ref{eq_corePotential}) the electrostatic potential
  $\Phi_e \propto {\cal E}_{hom}(R_1) \cdot R_1$ scales linearly
  with the dot radius and electric field strength. This field also depends
  on the dot size, but other than in a plain capacitor it scales 
  roughly with ${\cal E}_{hom} \propto U(R_1) \cdot R_1^{-0.4}$. The reason for
  this scaling behavior is mainly the fact that the tip-crystal distance
  is kept constant while scaling the crystal size. Last but not least the
  applied tip-substrate voltage depends on the energy needed to add
  a further electron to the crystal which is again a function of the
  dot radius. In an infinite potential well the single-particle energy
  levels scale with $R_1^{-2}$ but due to the finiteness of the 
  studied potential well and the fact that the effective mass increases 
  with increasing energy, leads to a scaling of roughly
  $U \propto R_1^{-1}$. Putting all together we find that the electrostatic
  potential scales with $\Phi_e \propto R_1^{-0.4}$. Therefore the LDOS
  symmetry changes from spherical to torus-like with increasing the
  crystal radius, since Stark energy becomes in bigger crystals more 
  important than exchange energy.

  Especially the scaling of the electrostatic potential is not straight forward
  and therefore it is necessary to check this result by a full calculation.
  To this end, the charging energy needed to add the fifth electron to the QD 
  has to be calculated in a first step. As shown in equation
  (\ref{Eq_TotalEnergy}) this charging energy is the energy difference between
  the 5- and 4-electron ground-states. In a second step the tip-substrate
  voltage needed to open this 5-electron channel has to be calculated. As shown
  in equation (\ref{Eq_peakpos}) this voltage is found by multiplying the 
  charging energy with the pre-factor $\gamma$, obtained from the
  electrostatic potential drop along z-direction (see Section
  \ref{sec:model}). Knowing the applied tip-substrate voltage, the electric 
  field strength in the QD can in a last step be calculated by equation 
  (\ref{eq_corePotential}). Since the charging energy calculated in the first step
  also depends on the electric field, the whole cycle is repeated until
  self-consistency is obtained. Now knowing the electric field strength we can
  go back into Fig. \ref{fig_GSCrossing} and determine the ground-state 
  configuration for the considered QD size and therefore determine the LDOS 
  symmetry.
  \begin{figure}[h]
    \begin{center}
      \includegraphics[scale=0.5]{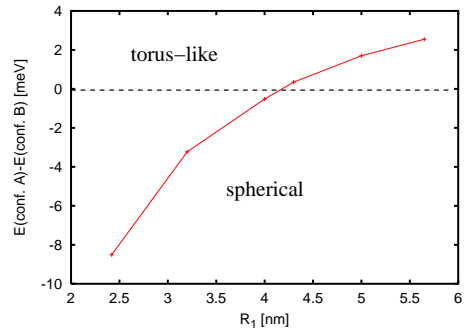}
    \end{center}
    \caption{The difference in energy between the 5-electron ground-state and
             first excited state is shown as a function of nanocrystal
             radius. For an energy difference greater than zero, configuration
             B shown in Fig. \ref{fig_configuration}c is the new ground-state
             which has a torus-like LDOS symmetry.}
    \label{fig_radiusscaling}
  \end{figure}
  We have done this calculation for six InAs crystals with radii 
  between about 2 and 6 nm and plotted the energy difference between 
  configuration A and B versus the dot radius in Fig. 
  \ref{fig_radiusscaling}. In this plot a number smaller than zero
  corresponds to a spherical and a number bigger than zero to a
  torus-like ground-state symmetry. As already predicted by the
  scaling considerations we found that the LDOS symmetry 
  changes from spherical to torus-like by increasing the crystal radius. 
  Using the material constants for InAs we found this transition to
  happen at a dot radius of about 4 nm.

\section{Conclusion}
  We calculated the Stark effect on many-particle wave-functions
  in InAs nanocrystals charged with up to five electrons. Stark effect
  and Coulomb interactions have been fully included by an exact
  diagonalization procedure. We found that the LDOS, due to the Stark-induced 
  degeneracy lifting of the $1P_e$ states, of the 3- and 4-electron 
  ground-states are torus-like in shape. For 5 electrons however, a
  competition between Stark energy and exchange energy leads to a
  ground-state crossing with increasing field. Since
  the electric field strength can not be directly controlled in
  the usual experiments, we
  studied the crystal size dependence of the 5-electron ground-state
  symmetry and found a transition from spherical to torus-like
  by increasing the dot radius. This transition should be observable
  in a wave-function mapping experiment.

\section{Acknowledgments}
  The authors gratefully acknowledge valuable discussions with
  Markus Morgenstern and Theophilos Maltezopoulos.
  This work was supported by the DFG through SFB 1641 and GrK 32048.

\section*{References}
\bibliographystyle{prsty}
\bibliography{biblio}

\begin{thebibliography}{10}

\bibitem{millo:06:2001}
O. Millo, D. Katz, Y. Cao, and U. Banin, Phys. Rev. Lett. {\bf 86},  5751
  (2001).

\bibitem{grandidier:07:2000}
B. Grandidier {\it et~al.}, Phys. Rev. Lett. {\bf 85},  1068  (2000).

\bibitem{vdovin:10:2000}
E.~E. Vdovin {\it et~al.}, Science {\bf 290},  122  (2000).

\bibitem{banin:08:1999}
U. Banin, Y. Cao, D. Katz, and O. Millo, Letters to Nature {\bf 400},  542
  (1999).

\bibitem{alperson:09:1999}
B. Alperson {\it et~al.}, Applied Physics Letters {\bf 75},  1751  (1999).

\bibitem{tews:02:2002}
M. Tews and D. Pfannkuche, Phys. Rev. B. {\bf 65},  073307  (2002).

\bibitem{katz:05:2001}
D. Katz, O. Millo, S.-H. Kan, and U. Banin, Applied Physics Letters {\bf 79},
  117  (2001).

\bibitem{millo:06:2000}
O. Millo, D. Katz, Y. Cao, and U. Banin, Phys. Rev. B. {\bf 61},  16773
  (2000).

\bibitem{bakkers:09:2000}
E.~P. Backers and D. Vanmaekelbergh, Phys. Rev. B. {\bf 62},  7743  (2000).

\bibitem{brus:05:1984}
L. Brus, Journal of Chemical Physics {\bf 80},  4403  (1984).

\bibitem{burt:03:1992}
M.~G. Burt, J. Phys.: Condens. Matter {\bf 4},  6651  (1992).

\bibitem{wiesendanger:1994}
R. Wiesendanger, {\em Scanning Probe Microscopy and Spectroscopy}, 1 ed.
  (Cambridge University Press, Cambridge, 1994).

\bibitem{griffiths_edyn:1998}
D.~J. Griffiths, {\em Introduction to Electrodynamics} (Prentice, Hall, 1998).

\bibitem{madelung:1996}
O. Madelung, {\em Semiconductors-Basic Data}, 2nd revised ed. (Springer,
  Berlin, 1996).

\bibitem{kim:06:2001}
B.~S. Kim, M.~A. Islam, L.~E. Brus, and I.~P. Herman, Journal of Applied
  Physics {\bf 89},  8127  (2001).

\bibitem{goldoni:06:2000}
G. Goldoni {\it et~al.}, Physica E {\bf 6},  482  (2000).

\bibitem{hallam:01:1996}
L.~D. Hallam, J. Weis, and P. Maksym, Phys. Rev. B. {\bf 53},  1452  (1996).

\bibitem{schiff:1993}
L.~I. Schiff, {\em Quantum Mechanics} (Springer, Berlin, 1993).

\bibitem{williamson:06:1999}
A.~J. Williamson and A. Zunger, Phys. Rev. B. {\bf 59},  15819  (1999).

\end{thebibliography}

\end{document}